# FEMSIM+HRMC: Simulation of and Structural Refinement using Fluctuation Electron Microscopy for Amorphous Materials

Jason J. Maldonis[1], Jinwoo Hwang[1], and Paul M. Voyles[1]

[1]Department of Materials Science and Engineering, University of Wisconsin, Madison, Madison, WI, 53706, USA

Abstract

FEMSIM, a Fortran code, has been developed to simulate the fluctuation electron microscopy signal, the variance, $V(k)$, from a model atomic structure. FEMSIM has been incorporated into a hybrid-reverse Monte Carlo code that combines an embedded atom or Finnis-Sinclair potential with the deviation between simulated and experimental $V(k)$ data to refine an atomic model with structure constrained by both the potential and experimental data. The resulting models have experimentally-derived medium-range order.

**Program Summary**

*Program Title*: FEMSIM + HRMC
*Catalog Identifier:*
*Program summary URL*:
*Program obtainable from:* University of Madison, WI
*Licensing provisions*: MIT Open Source
*No. of lines in distributed program, including test data, etc.*:
*No. of bytes in distributed program, including test data, etc.*:
*Distribution format*: tar.gz
*Programming language:* Intel FORTRAN 2000
*Computer*: Any computer with an Intel FORTRAN compiler that supports MPI
*Operating System:* Linux
*RAM:* For system sizes of approximately 1,000 atoms, less than 8 GB RAM is required
*Classification:*
*Nature of problem:* Simulation of fluctuation microscopy experimental data; atomic modeling using empirical potentials and experimental data
*Solution method:* Monte Carlo and Simulated Annealing
*Additional comments:* An Intel Fortran compiler mpif90 (OpenMPI) is recommended
*Running time:* FEMSIM alone takes < 1 minute on a single core for a model with approximately 1,000 atoms. HRMC structure refinement for the same model size converges in ~6000 CPU hours.

# 1. Introduction

Structure determination algorithms provide a rich set of tools for generating complex structures from energetic or experimental constraints. One of the most general and computationally feasible algorithms is the Metropolis Monte Carlo[1] (MMC) optimization framework. When applied in an effort to determine the structure of a material, MMC, coupled with simulated annealing[2] (SA), identifies preferred structure configurations by stochastically minimizing the system's potential energy, $E$. Given an initial structure and temperature, atoms or molecules are randomly moved to abruptly change the system's configuration, independent of energetic forces. The viability of the particle movement is determined by the relative change in $E$. Movements that decrease $E$ are accepted, while movements that increase $E$ are accepted probabilistically based on the system temperature[3].

Reverse Monte Carlo (RMC)[4–7] modifies the MMC + SA optimization by replacing the energy objective function, $O$, in MMC with an objective function that quantifies a fit to experimental data. By minimizing $O = \chi^2 = \sum_i (f_{exp}(x_i) - f_{sim}(x_i))/\sigma_i^2$, where $f(x)$ represents a set of numerical data and $\sigma_i^2$ is the variance of that data, RMC guarantees the final structure will reproduce the experiment, subject to the accuracy of the simulated experiment. However, the final structure generated is only guaranteed to reproduce the experimental results that are constrained by the objective function, and are therefore often energetically or otherwise unrealistic[8,9]. For example, in 2012, Treacy and Gibson[9] found that RMC simulations constrained against pair correlation or structure factor data from x-ray diffraction experiments of amorphous silicon failed to generate structures that matched data from fluctuation electron microscopy (FEM) experiments. In this case, the amorphous silicon models generated by RMC were meaningfully underconstrained. Additional constraints can be introduced into the objective function—such as bond angle and coordination number constraints in the case of amorphous carbon[3,10]—to mitigate this problem. In principle, adding additional constraints improves the simulation accuracy. However, the complexity of constructing an unbiased objective function increases with the number of constraints, and finding a configuration of atoms that simultaneously minimizes all of the constraints in the cost function becomes more difficult.

Hybrid reverse Monte Carlo (HRMC) is a variation of RMC that includes the energetic constraint of MMC. First introduced by Opletal et al.[11], HRMC defines a new objective function, $O = E + \alpha\chi^2$ where $\alpha$ is a weighting factor. HRMC is useful when the potential used for the energy calculation does not reproduce some experimental data, and the experimental data alone does not reproduce some well-defined aspect of the structure such as chemical ordering. Including both constraints in the objective function allows the HRMC simulation to optimize against all available information. By changing the random number generator seed or initial configuration of a simulation, multiple unique structures can be generated. These structures provide the necessary statistical data to characterize disordered materials.

This work focuses primarily on an implementation of HRMC to study disordered materials, especially metallic glasses, by incorporating experimental data from fluctuation electron microscopy (FEM) experiments[12]. FEM identifies structural heterogeneity in materials at the nm length scale. Hundreds of electron diffraction patterns are collected from a thin sample, and the variance of the images—after azimuthal averaging—is calculated by

$$V(k, R) = \frac{\langle I^2(k, R, \mathbf{r})\rangle_r}{\langle I(k, R, \mathbf{r})\rangle_r^2} - 1 \qquad (1)$$

where $I$ is the diffracted intensity, $k$ is the scattering vector magnitude, $R$ is the experimental probe resolution, and $\langle \rangle_r$ indicates averaging over position on the sample, $\mathbf{r}$. Formally, $V$ depends on the three- and four-atom distribution functions, but the inversion problem to obtain the correlation functions from the data is not solved[12]. Qualitatively, heterogeneous nanoscale structure increases the spatial fluctuations in nanodiffraction, increasing $V$. The $k$ position of peaks in $V(k)$ carries information about the internal structure of diffracting regions. Interpretation of $V(k)$ is rarely intuitive or straightforward, so we simulate FEM within HRMC to generate atomic models that match the experiment, and then analyze the models. We define

$$\chi^2 = \sum_i \frac{\left(\beta\, V_{exp}(k_i) - V_{sim}(k_i)\right)^2}{\sigma_i^2} \qquad (2)$$

where the summation over $i$ represents the discretized $k$ points at which the experiment was conducted, $\sigma_i^2$ is the experimental error at point $k_i$, and $\beta = 3 t_{exp}/t_{sim}$ is a scaling factor accounting for the difference in experimental and simulated sample thickness, $t_{exp}$ and $t_{sim}$, respectively. The factor of 3 is an approximate correction for the approximations involved in calculating $V(k)$ described below.

By including FEM data in the objective function of HRMC, the structure of the material is constrained at short-range (primarily nearest neighbor) by the potential and at the medium-range (nanometer) scale by the experimental data. Here, we provide an implementation of an FEM simulation, which we call FEMSIM, based on work by Dash et al.[13] and incorporate this algorithm into an HRMC framework. Section 2 discusses implementation details of the algorithms. Section 3 illustrates the scalability of the code for use on multi-core computers via MPI parallelization. Section 4 details the necessary inputs and outputs of the simulations, and Section 5 provides suggestions for setting up an accurate and computationally feasible simulation. Section 6 concludes with examples and applications from the literature.

## 2. FEMSIM and HRMC

### 2.1 FEMSIM

The program begins by setting up a message passing interface (MPI) using standard Fortran MPI functions. The program then initializes and allocates memory for data structures using information contained in a parameter file (see Section 4 for details). Included in this data structure is a linked cell list[14,15] (called a "hutch" in the code comments) designed to improve the computation time of both the potential energy and $V(k)$ calculations. After initialization, the FEMSIM calculation is performed, and $V(k)$ is saved to disk.

FEMSIM simulates the FEM experiment: given an atomic model, it computes an approximation of $V(k)$. FEMSIM starts by rotating the input model ~ 200 times with angles that uniformly sample the unit sphere. The 2D projections of these models along the z-axis are unique. These rotations assume that the long-range structure of the sample is isotropic. Given that the structure is isotropic, orientation averaging in the simulation gives the same result as spatial averaging in the experiment. FEMSIM calculates $V(k)$ via the Dash et. al method[13]. A local pair distribution function about position $r$ defined by

$$g_{2A}(r',r,Q) = \sum_j \sum_i A(2\pi Q|r-r_j|)A(2\pi Q|r-r_i|)\delta(|r_i - r_j| - r') \quad (3)$$

is used to calculate the intensity $I(r,k,Q)$ by numeric integration of $g_{2A}(r',r,Q) \times J_0(2\pi kr')$ over $r'$ for the desired values of $k$ and $Q = 0.61/R$. In Eq (3), $A$ is the Airy function, $A(x) = \frac{J_1(x)}{x}$, and $J_0$ and $J_1$ are the zeroth- and first-order Bessel functions. In FEMSIM, $Q$ is determined by the probe resolution). $V(k)$ is then calculated via Eq (1), where each $I(r,k)$ is produced by a section (designated by $r$) of a unique rotation of the model. The sections of the model are $(\sqrt{2}R) \times (\sqrt{2}R)$ in length and width (for each rotation), and extend through the model. This method is an approximation of $V(k)$. Simulations can be brought into better agreement with experiments bymultiplying $V_{sim}(k)$ by 1/3, which is a typical correction for the "Stobbs factor" between electron microscopy simulations and experiments[12,13,16]. $I(k)$ calculations are independent, so they are evenly distributed over the number of available cores.

In the FEM experiment, the electron probe is circular in shape. However, the atomic model in the simulation must obey periodic boundary conditions and therefore cannot be spherical. We approximate the shape of the experimental probe as a square with the circular probe of radius $R$ inscribed inside, as shown in Figure 1. This restricts the side length of the cubic model to be $l = \sqrt{2}Rn$ where $n$ is an integer.

*2.2 HRMC*

HRMC incorporates FEMSIM into an optimization scheme by calculating $V_{sim}(k)$ and $\chi^2$ at every step of the optimization. $\chi^2$ is then used in the objective function, $O = E + \alpha\chi^2$, which is minimized via MMC + SA. After $V_{sim}(k)$ is calculated from the initial model configuration, the potential energy of the initial configuration is computed, and the initial values of $E$ and $V(k)$ are stored. The program then enters a loop, which begins by moving a randomly chosen atom in a random direction by a random distance and propagating that change into the rotated models. $E$ and $V(k)$ are recalculated, taking care to only update values that are within the relevant region of the moved atom in order to avoid unnecessary computation. The cost function is then reevaluated, and the atom move is accepted or rejected based on the MMC algorithm. If the cost function decreased with respect to the previously accepted move, the atom move is accepted. If the value of the cost function increased, a random number $rand \in (0,1]$ is generated and the move is accepted if $rand < \exp\left[-\frac{\Delta CF}{kT}\right]$ where $k$ is Boltzmann's constant and $T$ is the temperature of the system. (Since the cost function is not an energy, $T$ is not a physical temperature, but it plays the same role in the MMC part of the algorithm.) If the atom move is rejected, all values generated during this step are discarded and the atom is reset to its original position in all models. Every $N$ steps (predefined by the user) the temperature and maximum distance an atom can move are decreased via an inverse power law, analogous to simulated annealing. This process is repeated until a pre-defined number of steps has been executed, or until the user manually stops the program. It is up to the user to determine when the cost function has converged, which indicates the end of the simulation.

## 3. Scalability of FEMSIM in HRMC

The user may wish to run HRMC on a multicore machine. To this end, FEMSIM has been parallelized to work with MPI. FEMSIM is parallelized at the level of the rotated models, and calculations of $I(r, k)$ for a given rotated model are distributed evenly across the number of cores. As a result, the code does not scale past one core per rotation per pixel. The results in this section were gathered from simulations with $R = 2$ nm, 1250 atoms, one pixel, and 211 total rotation angles and run on architecture such as Intel(R) Xeon(R) CPU E5-2697 v3 @ 2.60GHz) unless specified otherwise.

Figure 2a shows the decrease in the average time per step for an HRMC simulation, averaged over the first 100 atom moves, as a function of the number of cores. Ideally, doubling the number of cores will halve the wall-clock time of the simulation. This is not true in practice, however, because the entirety of the code is not parallelizable. Instead, the scalability follows Amdahl's law[17]. Figure 2b shows the speedup of HRMC under the same conditions as above, as a function of the number of cores used. Using the fit to Amdahl's law, we find that 99.5% of HRMC's serial computation time is parallelized using the input parameters above. This implies that the calculation of $I(r, k)$ is the dominant source of computation time in HRMC, and hence we have not parallelized other portions of the code. Unfortunately, increasing the number of atoms decreases the speedup because proportionally more computation time is added to the serial portion of the code. We found that shared-memory parallelization (via OpenMP) within calculations of $I(k)$ did not provide an increase in speed.

Running on 64 cores, a simulation with the parameters above requires approximately 6000 CPU hours (about four days) to converge with a reasonable cooling rate. Increasing the model size to $l = 2R = 4\ nm$ (four pixels and 8x more atoms) requires approximately two months of computation time, or 200,000 CPU hours. The FEMSIM algorithm scales quadratically with the number of atoms, and the number of steps until convergence scales linearly with the number of atoms, leading to a cubic dependence of the scaling on the number of atoms as the system size increases.

## 4. Input and Output Parameters and Files

*4.1 Compiling*

All code is written in FORTAN and can be compiled using an Intel Fortran 90 compiler that supports MPI (for example, Intel's mpif90). Compilation should be done using GNU Make and the makefile provided in the *src* directory. A make target is required, and can be either *hrmc* or *femsim*. An optional keyword *debug* will compile the code with debugging enabled. For example, *make hrmc debug –C src* (run from the root directory) will create an executable with the name *hrmc*, compiled without optimizations and with debugging enabled. Once compiled, the code can be run with *./executable basename paramfile.in* where *executable* is either *hrmc* or *femsim*, *basename* specifies a

keyword that is used as part of the output filenames to prevent output files from being overwritten when multiple simulations are run in the same directory, and *paramfile.in* is the name of a parameter file.

*4.2 Input files*

The code is contained in the following files:
- eam.f90 and eam_fs.f90 define the EAM/alloy and EAM/Finnis-Sinclair potential energy functionality, respectively
    - The user will need to specify which functional form should be used by including either eam.f90 or eam_fs.f90 within the source files in the makefile.
- fem.f90 contains the code relevant to the FEMSIM calculation, including its usage within HRMC
- globals.f90 defines a few global variables (e.g. the MPI variables) for the entirety of the code
- model.f90 defines the code relevant to handling the atomic model
- read_inputs.f90 parses the parameter file and sets relevant variables
- rmc_functions.f90 defines functions necessary for the HRMC looping schema
- hrmc.f90 contains the core code that runs HRMC or FEMSIM, depending on the make command
- scattering_factors.f90 defines the electron scattering factors[18] used within the FEMSIM calculation

Additional input files for FEMSIM include:
- an atomic model file in XYZ format
    - There is one additional formatting requirement: The comment line must start with three floats (separated by a space) indicating the size in Angstroms of the atomic model. The atomic model is required to be cubic.
- a file containing a list of $k$ points for which $V_{sim}(k)$ will be evaluated
    - The first row is comment line, followed by any number of $k$ values, each on a separate line.
- a submit file for submitting the code to a cluster (optional)

Additional input files for HRMC include:
- an atomic model file in XYZ format
    - There is one additional formatting requirement: The comment line must start with three floats (separated by a space) indicating the size in Angstroms of the atomic model. The atomic model is required to be cubic.
- an experimental data file containing $V(k)$ information
    - The first row is a comment line, followed by any number of rows of data points.

o  The data is space-delimited and must be three columns long: 1) $k$, at which $V_{sim}(k)$ will be evaluated; 2) $V_{exp}(k)$ for comparison to $V_{sim}(k)$; and 3) $\sigma_i^2(k)$.
- an EAM/alloy or EAM/Finnis-Sinclair potential file in standard format, with one float per line
    o  the helper file `utils/reformat_potential.py` may be helpful for putting a single float on each line
- a submit file for submitting the code to a cluster (optional)
    o  An example using the SLURM scheduling system is included.

The parameter file must have the following format on each line:
1. comment line, ignored by the program
2. input atomic model filename
3. input experimental data filename
4. $Q = 0.61/R$ (defined by the Rayleigh resolution criterion)
5. number of rotations: $n\phi$, $n\psi$, $n\theta$
    a. $\phi, \psi$, and $\theta$ follow the Goldstein convention[19], and $n\phi, n\psi$, and $n\theta$ represent the number of angles about their respective axes.
    b. The angles are chosen such that the unit sphere is uniformly sampled.
    c. $n\phi$ is the number of angles about the z-axis and should be set to 1 for all practical purposes.
    d. $\phi \in [0, 2\pi) = \pi$, $\psi \in [0, 2\pi)$, $\theta \in [0, \pi)$
    e. In the example, $n\phi = 1, n\psi = 40, n\theta = 20$

The following lines can be omitted for FEMSIM but are required for HRMC:
6. thickness scaling factor: $\beta = 3 t_{exp}/t_{sim}$
7. input potential filename
8. *starting_step, ending_step*
    a. *starting_step* is the step number at which the simulation will start and is useful for continuations of a previous simulation.
    b. *ending_step* is the step number at which the simulation will terminate and is most useful when a cluster's queuing system has a maximum job duration.
9. *temperature, max_move, decrement*
    c. *temperature* is the temperature of the simulation at *starting_step* = 0.
    d. *max_move* is the maximum distance an atom is allowed to move at *starting_step* = 0.
    e. *decrement* is the number of steps that are run before *temperature* and *max_move* are decremented via an inverse power law.
10. random number generator seed (integer only)
11. weighting factor, $\alpha$, between $E$ and $\chi^2$
12. number of atom species in the atomic model
13. The next N lines are a list of N hard sphere cutoff distances, where N is the number of atom species. The column/row matrix has the same order (from left to right and top to bottom) as the atom species in the potential file, and values on each line must be separated by a space. When an atom is randomly moved in the HRMC algorithm, the move will be immediately retried without calculation of the energy or $\chi^2$ if moved atom is within the hard sphere cutoff distance of another atom. We suggest using the minimum interatomic distance from $g(r)$ data for the hard sphere cutoff values.

*4.3 Output files*

The *basename* keyword in the following names is a placeholder for a command line input (see Section 4.1) that prevents output files from being overwritten when multiple simulations are run in the same directory. In the case that HRMC is submitted to a cluster, *basename* can be replaced by the job ID.

- stdout displays step-by-step information, including whether an atom move was accepted or rejected and the values that pertain to the acceptance criterion.
    - Written to every step.
- stderr displays all error related information.
    - Written to when an error occurs.
- acceptance_rate_basename.txt contains a single comment line followed by columns of step numbers, in increments of 100, and acceptance rates averaged over the previous 100 steps.
    - Written to every 100 steps.
- chi_squared_basename.txt contains a comment line followed by columns of step number, $\chi^2$, and *E*.
    - Written to every time an atom move is accepted.
- model_update_basename_N.xyz contains the model file in XYZ format at step *N*.
    - Written once every 1000 steps.
- model_final_basename.xyz contains the model file in XYZ format at the end of the simulation.
    - Written once at the end of the simulation.
- vk_initial_basename.txt contains the simulated *V(k)* for the starting model.
    - Written once at the beginning of the simulation.
- vk_final_basename.txt contains the simulated *V(k)* for the final model.
    - Written once at the end of the simulation.

## 5. Simulation Setup

This section gives best practices for setting up FEMSIM and HRMC in a usable manner. The inputs for FEMSIM are relatively straightforward. All distances should have units of angstroms, the atomic model file must be cubic with $l = \sqrt{2}Rn$ where *n* is an integer and *R* is the experimental probe resolution, and the total number of rotation angles should be large ($\geq 200$).

The accuracy and time-to-convergence of the HRMC simulation is largely dependent on the input parameters, so users must take care when setting up the simulation. The initial atomic model must have the correct atomic density, and it is often more important for the density to agree with the potential than with experimental results. We encourage the user to equilibrate an atomic model with the correct composition via molecular dynamics in an NPT ensemble (at the temperature of the experiment). When the volume of the simulation box has equilibrated, the user can calculate the atomic density. An atomic model of any size can be input to the molecular dynamics simulation, so long as finite size effects are avoided. An atomic model with the correct box size for the HRMC simulation can then be generated, and a conjugate gradient energy minimization is encouraged before final submission to the HRMC simulation. The HRMC simulation requires the atomic model to be cubic, to obey periodic boundary conditions, and the length of an edge to be $l = \sqrt{2}Rn$.

Hard-sphere cutoff distances are necessary to avoid extremely poor atom moves and prevent unnecessary computation of $E$ and $V_{sim}(k)$. These values can be found from the minimum interatomic distance of partial pair correlation functions (calculated, for example, using RINGS[20]) from the equilibrated molecular dynamics model.

The weighting factor, $\alpha$, between $E$ and $\chi^2$ plays a crucial role in determining the potential energy of the final model and the fit to experiment. Higher values of $\alpha$ will result in a better fit to experimental data, often at the cost of a higher potential energy. Empirically, we find that $\alpha$ set so that $\Delta(\alpha E) \approx \Delta\chi^2$ is a good rule of thumb. However, the ratio of these values changes during the simulation and thus $\alpha$ is difficult to determine *a priori*. We have found that estimating $\alpha$ using $\alpha = \Delta(\alpha_0 E)/\Delta\chi^2$ (where $\alpha_0$ is the value of $\alpha$ used in a test simulation) for values of $\Delta(\alpha_0 E)$ and $\Delta\chi^2$ near the end of the simulation is most useful. A HRMC simulation with a fast decrease in temperature may be helpful to determine this value.

*max_move* is the maximum distance an atom can move in any coordinate direction. It is decreased via a power law function during the simulation to prevent the acceptance rate from becoming unreasonably low. For previous HRMC simulations, we used an initial value for *max_move* equal to ~ 75% of the average hard-sphere cutoff distance, which resulted in atoms moving 6-8 Å over the course of the simulation. If *max_move* is too small, it will limit the mobility of the atoms and significantly increase the number of necessary steps before convergence because many moves may result in inconsequential changes. If this value is too large it will result in lower acceptance rates as well as unnecessary computation because a random move that satisfies the hard-sphere cutoff distances will be unlikely, and thus the algorithm will retry the move many times.

Once *max_move* has been set, the user should determine the initial *temperature*. When an objective function with non-energetic data is used, such as in HRMC, the temperature of the system loses physical meaning and becomes fictitious. However, this value is critical in controlling the acceptance rate of atom moves and therefore the time-to-convergence of the simulation. As a rule of thumb, MMC should have an acceptance rate of ~ 50%. However, over the course of simulated annealing as the artificial temperature goes down, the fraction of moves that are accepted decreases significantly. To compensate for this, the starting temperature of the HRMC simulation should be set so that the acceptance rate is ~ 95%. The most straightforward way to set the temperature is by trial and error. The user may run a few thousand moves, examine the acceptance rate, and modify the temperature until the acceptance rate is ~ 95%. All other parameters, with the exception of *decrement*, should be set correctly before determining the starting temperature.

Once the initial temperature has been set, the user can determine *decrement*. *decrement* sets the interval (in units of steps) between decrements of *temperature* and

*max_move*. *decrement* should be large enough that the cost function has time to equilibrate, while being small enough to avoid unnecessary computation at equilibrium. If *decrement* is too large, the value of the cost function will plateau within each *decrement* window, which indicates that HRMC is performing unnecessary computation after equilibrium has been reached at a given artificial temperature. If *decrement* is too small, the model may not have time to equilibrate and the cost function will decrease continuously, but not to the low value possible with a better value of *decrement*.

After all parameters have been set, there are two additional quality checks that can be helpful. During initialization of FEMSIM, an *M x M* grid of numbers is printed to stdout. Each row and column represents a 2D projection through the z-axis of the initial model with a width and height of ~ 1 Å. The numbers in each column represent the number of atoms that have been found in the 2D projection that will be used to calculate $V(k)$. This grid should be relatively uniform, and it should be clear to the user whether the entire model is being captured within the FEMSIM calculation. Secondly, it can be helpful to run FEMSIM on a model that has been generated by HRMC with additional $k$ points added to the $V(k)$ calculation. If the original input $k$ data is too sparse, HRMC may generate a model that agrees well with the included data points, but does not correctly reproduce $V(k)$ between the provided sampled data.

Finally, if the user is submitting HRMC to a cluster with limited job duration and the simulation will require multiple sequential jobs before completion, a modifiable python helper script is provided to make this process easier (see *submits/slurm_submit.py* for a SLURM version).

## 6. Examples

We provide one example FEMSIM and two example HRMC simulations. Other use cases can be found in references [21–23].

*FEMSIM*

The parameter file *examples/femsim/paramters/femsim.in* in the submitted code repository provides input parameters to evaluate $V_{sim}(k)$ for a $Zr_{50}Cu_{45}Al_5$ metallic glass model with 1250 atoms and *R* = 2 nm (the model file can be found in *examples/femsim/models/Zr50Cu35Al15_t3_final.xyz*). This data was recently published as part of a study of glass forming ability of metallic glasses[21]. The correct values for this $V_{sim}(k)$ evaluation can be found in *examples/femsim/ouputs/vk_initial_femsim.txt* and are shown in Figure 3a.

*HRMC*

The parameter file *examples/hrmc/parameters/hrmc.in* in the submitted code repository provides input parameters to run a tiny HRMC simulation so the user can test their compilation. The parameter file runs 1000 HRMC steps using nine rotations on a $Zr_{50}Cu_{45}Al_5$ model with 156 atoms, *R* = 1 nm, and a random number seed of 10475. The outputs are recorded in *examples/hrmc/outputs/*. $V(k)$ for the initial model, the final model after 1000 steps, and the experimental data, as well as the initial and final models, are shown in Figure 3b. We expect this test simulation to take less than one CPU hour.

A different parameter file, *parameters/hrmc.in*, will perform a full refinement of a $Zr_{50}Cu_{35}Al_{15}$ atomic model. The resulting model should be structurally analogous to the example FEMSIM model above, and details can be found in reference [21]. In addition to the starting model file, experimental data and an EAM/alloy potential file[24] are provided. Submit files for a clusters using either a SLURM or PBS scheduler are included in *submits/*. We expect this full simulation to take ~ 6000 CPU hours.

## 7. Conclusions

We provided a program to calculate the fluctuation electron microscopy variance, $V(k)$, from an atomic model file. This code, called FEMSIM, is incorporated into a hybrid reverse Monte Carlo (HRMC) framework to perform reverse structure optimization against experimental $V(k)$ data and the system energy calculated from an empirical potential. Together, the experimental data and potential constrain the short- and medium-range order of the atomic model to produce a realistic configuration that both matches experimental data and is energetically reasonable. The input parameters for both FEMSIM and HRMC are explained, and examples are provided.

**Acknowledgements**
This work was supported by the National Science Foundation under Contract No. DMR-1332851. Initial development of the code was supported by NSF DMR-1205899 and CMMI-1232731.

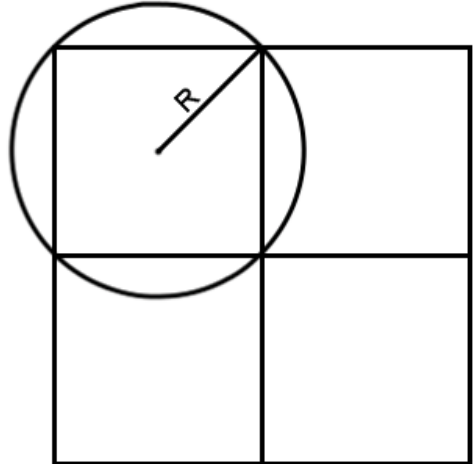

Figure 1: The experimental probe (circular) is reshaped into a square within FEMSIM.

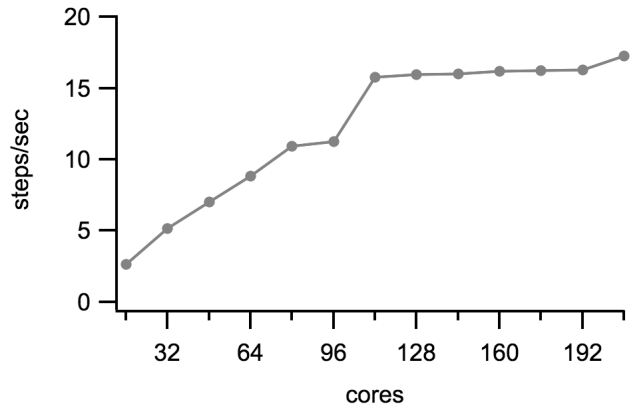

Figure 2(a): The number of HRMC steps computed per second as a function of the number of cores.

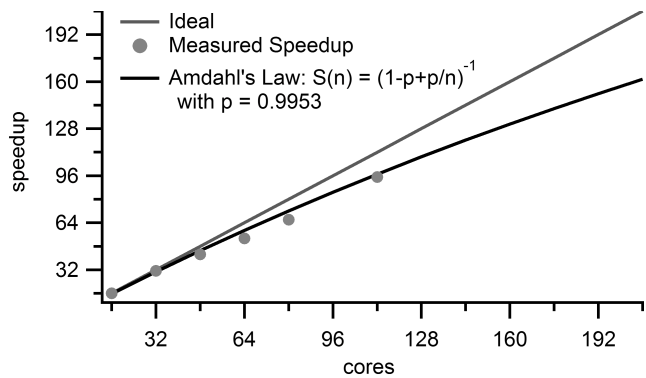

Figure 2(b): Speedup of HRMC as a function of the number of cores fit to Ahmdal's law and compared to the ideal case.

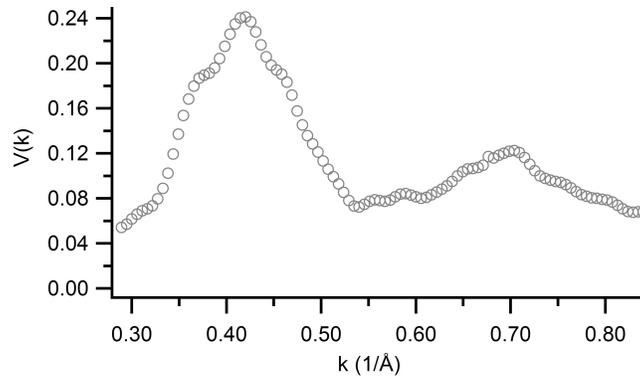

Figure 3(a): $V_{sim}(k)$ for a $Zr_{50}Cu_{45}Al_5$ metallic glass calculated via FEMSIM.

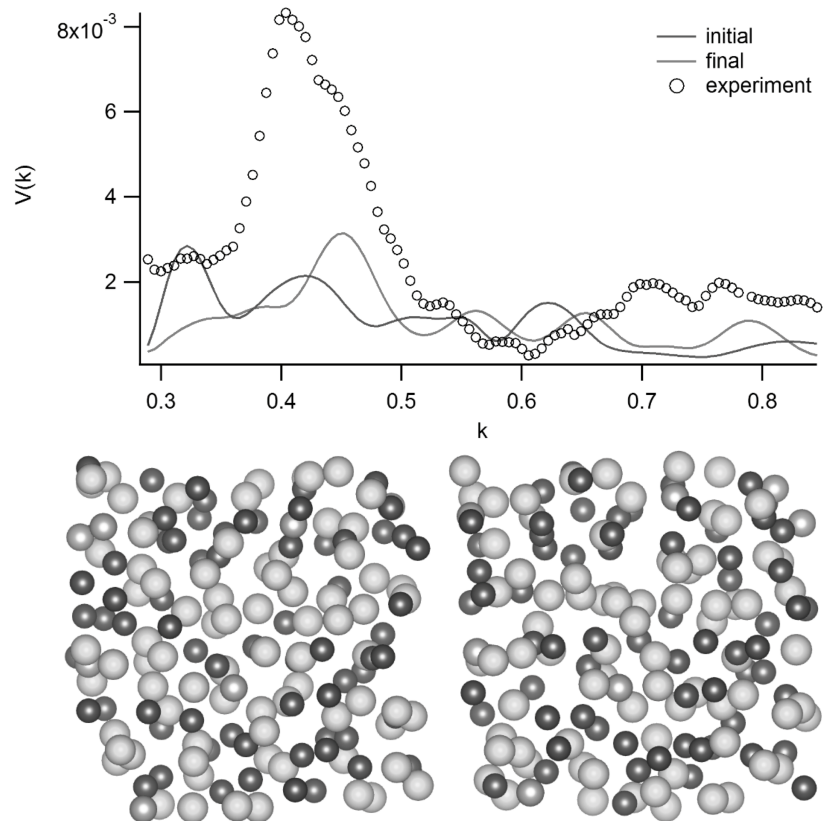

Figure 3(b): $V_{exp}(k)$ compared to $V_{sim}(k)$ at the beginning and end of a miniature HRMC simulation for a small $Zr_{50}Cu_{45}Al_5$ metallic glass along with the initial (left) and final (right) model structures.